\documentstyle[prl,aps,preprint]{revtex}
\let\ssection=\section
\renewcommand{\section}{\setcounter{equation}{0}\ssection}

\newcommand{\be}{\begin{equation}}
\newcommand{\ee}{\end{equation}}
\newcommand{\ba}{\begin{eqnarray}}
\newcommand{\ea}{\end{eqnarray}}
\newcommand{\bec}{\begin{center}}
\newcommand{\eec}{\end{center}}

\begin{document}
\draft
\title{On the chiral anomaly in non-Riemannian spacetimes} 

\author{Yuri N.\ Obukhov\footnote{Permanent address: Department of
Theoretical Physics, Moscow State University, 117234 Moscow,
Russia}\and Eckehard W.\ Mielke,
%$^{\diamond}$\footnote{Permanent address: Faculty of Science, 
%University of Kiel, Kiel, Germany}
Jan Budczies, and Friedrich W.\ Hehl}
\address{Institute for Theoretical Physics, University of
Cologne\\D-50923 K{\"o}ln, Germany\\
$^{\diamond}$Departamento de F\'{\i}sica,\\ 
Universidad Aut\'onoma Metropolitana--Iztapalapa,\\ 
P.O. Box 55-534, 09340 M\'exico D.F., Mexico\\ }

\maketitle
\bigskip
%\noindent{\it file anomaly4.tex, 1997-04-Feb}
\bigskip

\begin{abstract} 
The {\it translational} Chern-Simons type three-form {\it coframe}
$\wedge$ {\it torsion} on a Riemann-Cartan spacetime is related (by
differentiation) to the Nieh-Yan four-form. Following Chandia and 
Zanelli, two spaces with non-trivial translational Chern-Simons forms are 
discussed. We then demonstrate, firstly within the classical 
Einstein-Cartan-Dirac theory and secondly in the quantum heat kernel 
approach to the Dirac operator, how the Nieh-Yan form surfaces in both 
contexts, in contrast to what has been assumed previously.
\end{abstract}
\bigskip\bigskip
%\pacs{PACS no.: 04.50.+h; 04.20.Jb; 03.50.Kk}
\bigskip
%\pagebreak

\section{Introduction}

We will address a question within metric-affine gravity,
the gauge theory of the affine group $R^4\;{\rlap{$\subset$}\times}\;GL(4,R)$. 
In this framework, the linear groups $SL(3,R)$ \cite{Dothan} and
$SL(4,R)$ \cite{Neeman} play decisive roles for the representation of
matter fields. Very early, L. Biedenharn and his collaborators
\cite{Biedenharn} or students \cite{Si1} investigated the
half-integer representations (of the covering group) of the $SL(3,R)$.
These representations turned out to be important for metric-affine
gravity, see \cite{PRs}. We dedicate our article to the memory of
L. Biedenharn.

In \cite{ChandiaZ}, following the earlier paper \cite{MardonesZ}, it has
been proposed that, in a Riemann-Cartan spacetime, the translational
Chern-Simons term \cite{He11,PRs} may play a  role in the 
chiral anomaly. However, in the context of this derivation, a 
{\it massless} Dirac field is used which is, in addition to the 
Poincar\'e invariance, also invariant under scale transformations. 
Therefore the appropriate spacetime arena is a Weyl-Cartan space. This 
space has an additional Weyl covector $Q$ which is expected to occur 
in the corresponding anomaly term.

In this paper we analyze geometrical and physical models which provide
some further evidence in support of possible ``topological'' manifestations 
of torsion and nonmetricity on the classical and the quantum level.  

%----------------------------------------------------------------------
\section{Topological invariants in four dimensions}

In metric-affine gravity, the $GL(4,R)$ Chern-Simons form reads (see, e.g., 
ref. \cite{He11})
\begin{equation}
C_{\rm RR}:=-{1\over 2}\left(\Gamma_\alpha{}^\beta\wedge d\Gamma_\beta{}^\alpha
- {2\over 3}\Gamma_\alpha{}^\beta\wedge\Gamma_\beta{}^\gamma\wedge
\Gamma_\gamma{}^\alpha\right),
\end{equation}
where $\Gamma_\alpha{}^\beta$ is the linear connection. The usual Pontrjagin 
(Chern) topological invariant is the exterior derivative of it,
\begin{equation}
-{1\over 2}R_\alpha{}^\beta\wedge R_\beta{}^\alpha = dC_{\rm RR}.\label{pont}
\end{equation}
It is straightforward to prove that the
cohomology class of the Pontrjagin four-form (\ref{pont}) is independent of
the connection. In the metric-affine space, we can perform a decomposition 
of all geometrical objects in Riemannian and post-Riemannian 
contributions. Denote the {\it distortion} one-form by $N_\alpha{}^\beta$.
It describes the difference between the general linear connection and the 
Levi-Civita (Riemannian) connection $\widetilde{\Gamma}_\alpha{}^\beta$:
\begin{equation}
\Gamma_\alpha{}^\beta = \widetilde{\Gamma}_\alpha{}^\beta + 
N_\alpha{}^\beta.\label{gamdecMAG}
\end{equation}
%The tilde denotes the purely Riemannian connection.
According to (\ref{gamdecMAG}), the curvature decomposes as follows:
\begin{equation}
R_\alpha{}^\beta = \widetilde{R}_\alpha{}^\beta + 
\widetilde{D}N_\alpha{}^\beta - N_\alpha{}^\lambda\wedge N_\lambda{}^\beta.
\end{equation}
Accordingly we find
\begin{equation}
R_\alpha{}^\beta\wedge R_\beta{}^\alpha =
\widetilde{R}_\alpha{}^\beta\wedge \widetilde{R}_\beta{}^\alpha
+ d\left[N_\alpha{}^\beta\wedge\left(2\widetilde{R}_\beta{}^\alpha
+ \widetilde{D}N_\beta{}^\alpha - {2\over 3}N_\beta{}^\lambda\wedge 
N_\lambda{}^\alpha\right)\right].
\end{equation}
Thus the difference of the total Pontrjagin form and the purely Riemannian
one is an exact form which proves their topological equivalence, see 
\cite{Mi6}. Incidentally, {\it deformed} Euler and Pontrjagin forms are 
discussed in \cite{Polonica} in the context of establishing the dynamical 
scheme for Poincar\'e gauge gravity. 

Direct calculations shows that the axial anomaly is expressed in terms of
the Riemannian Pontrjagin form. However, it is well known that there
exists an additional topological invariant in four dimensions: The `Nieh--Yan 
four-form' \cite{NY} which is defined as the exterior derivative of the 
translational Cherm-Simons term 
\begin{equation}
C_{\rm TT}:={1\over{2\ell^2}}
\vartheta^\alpha\wedge T_\alpha, 
\end{equation}
see \cite{He11}. In a generic metric--affine spacetime, the Nieh--Yan 
form reads:
\begin{equation}
\label{NiehYan} 
dC_{\rm TT} = {1\over{2\ell^2}} \left(T^{\alpha}\wedge
T_\alpha+R_{\alpha\beta}\wedge\vartheta^\alpha\wedge\vartheta^\beta
-Q_{\alpha\beta}\wedge\vartheta^\alpha\wedge T^\beta\right),
\end{equation}
where $\ell$ is the Planck length. In a Hamiltonian formulation \`a la
Ashtekar, this serves as a generating function for self-dual or chiral 
variables in gravity \cite{Mie92}.

Consider the first Bianchi identity, $DT^\alpha=R_\beta{}^\alpha\wedge
\vartheta^\beta$. Its third irreducible piece supplies only one equation
\cite{PRs},
\begin{eqnarray}
^{(3)}R_{\alpha\beta}\wedge\vartheta^\alpha\wedge\vartheta^\beta &=&
d\left(g_{\alpha\beta}\vartheta^\alpha\wedge T^\beta\right) + \left(
Q_{\alpha\beta}\wedge\vartheta^\beta - T_\alpha\right)\wedge T^\alpha
\nonumber\\
&=& d\left(\vartheta_\alpha\wedge\,^{(3)}T^\alpha\right) + \left(
{}^{(2)}Q_{\alpha\beta}\wedge\vartheta^\beta - {}^{(1)}T_\alpha\right)
\wedge^{(1)}T^\alpha \nonumber\\
&&+ \left({}^{(3)}Q_{\alpha\beta}\wedge\vartheta^\beta
+ {}^{(4)}Q_{\alpha\beta}\wedge\vartheta^\beta
-2\,^{(2)}T_\alpha\right)\wedge \,^{(3)}T^\alpha .\label{b1b} 
\end{eqnarray}
Here the irreducible pieces are denoted by superscripts enclosed by
parentheses. Specifically, the axial torsion is described by the third
irreducible part, $^{(3)}T^\alpha ={1\over 3}^*(\vartheta^\alpha\wedge A)$,
where the {\it axial} one-form is defined by $A:=-^*(\vartheta^\alpha
\wedge T_\alpha)$. Note that $^{(3)}R_{\alpha\beta}={1\over 2}
R_{[\alpha\beta\gamma\delta]}\,\vartheta^\gamma\wedge\vartheta^\delta$, i.e.,
its components correspond to the totally antisymmetric part of the curvature.
The expression in the last line of (\ref{b1b}) is typical 
for the post-Riemannian decompositions,
\begin{equation}
{}^{(3)}Q_{\alpha\beta}\wedge\vartheta^\beta
+ {}^{(4)}Q_{\alpha\beta}\wedge\vartheta^\beta -2\,^{(2)}T_\alpha 
= {1\over 3}\vartheta_\alpha\wedge(\Lambda - 3Q - 2T),
\end{equation}
where
\begin{equation}
\Lambda:=(e^\beta\rfloor {\nearrow\!\!\!\!\!\!\!Q}_{\alpha\beta})
\vartheta^\alpha,\quad {\nearrow\!\!\!\!\!\!\!Q}_{\alpha\beta}:=
Q_{\alpha\beta}- Qg_{\alpha\beta}, \quad Q :={1\over 4}Q_\alpha{}^\alpha,
\quad T:=e_\alpha\rfloor T^\alpha.
\end{equation}

If the torsion is totally antisymmetric, $T^\alpha={}^{(3)}T^\alpha$ (pure
``axial'' torsion), and the nonmetricity is purely Weyl, i.e., 
${\nearrow\!\!\!\!\!\!\!Q}_{\alpha\beta}=0$, then, in this
Weyl-Cartan spacetime, we have 
\begin{equation}
^{(3)}R_{\alpha\beta}\wedge\vartheta^\alpha\wedge\vartheta^\beta=
(d + Q)\left(\vartheta_\alpha\wedge\,^{(3)} T^\alpha\right)
\end{equation} 
or
\begin{equation}
{1\over 2}R_{[\alpha\beta\gamma\delta]}\,\vartheta^\alpha\wedge\vartheta^\beta
\wedge\vartheta^\gamma\wedge\vartheta^\delta=(d + Q)\,{}^\ast A\,.
\end{equation}
The proof of (\ref{b1b}) goes along the lines mentioned in the context 
of \cite{PRs} Eqs.(B.2.19) and (B.5.14).

%----------------------------------------------------------------------
\section{Spacetime with nontrivial Nieh-Yan four--form}

Consider the $SO(4)$-invariant Riemannian metric (with Euclidean signature), 
\begin{equation}
g=h^2\,dr^2 + f^2\left[d\psi^2 + \sin^2\psi\left(d\theta^2 +
\sin^2\theta d\phi^2\right)\right],\label{ds2}
\end{equation}
where $h=h(r), f=f(r)$, and $(r,\psi,\theta,\phi)$ are the standard
hyperspherical coordinates which parameterize the unit three-sphere $S^3$,
\begin{equation}
0\leq\phi < 2\pi,\quad\quad 0\leq\theta <\pi,\quad\quad 
0\leq\psi < \pi.
\end{equation}

In order to describe the metric-affine spacetime, we need to specify the
gravitational potentials ($g_{\alpha\beta}, \vartheta^\alpha, 
\Gamma_\alpha{}^\beta$). We choose these fields as follows: the metric as
\begin{equation}
g_{\alpha\beta}=\delta_{\alpha\beta},\label{metr}
\end{equation}
the coframe as the ``square root'' of (\ref{ds2}),
\begin{equation}
\vartheta^{\hat{4}}=hdr,\quad \vartheta^{\hat{1}}=fd\psi,\quad
\vartheta^{\hat{2}}=f\sin\psi d\theta,\quad 
\vartheta^{\hat{3}}=f\sin\psi\sin\theta d\phi,\label{cofr}
\end{equation}
and the linear connection eventually as
\begin{eqnarray}
\Gamma_{\hat{1}}{}^{\hat{2}}&=&-\Gamma_{\hat{2}}{}^{\hat{1}}=
\cos\psi\, d\theta - \sin\psi\sin\theta\, d\phi,\label{gam1}\\
\Gamma_{\hat{3}}{}^{\hat{1}}&=&-\Gamma_{\hat{1}}{}^{\hat{3}}=
-\sin\psi\, d\theta -\cos\psi\sin\theta\, d\phi,\label{gam2}\\
\Gamma_{\hat{2}}{}^{\hat{3}}&=&-\Gamma_{\hat{3}}{}^{\hat{2}}=
-d\psi + \cos\theta\, d\phi.\label{gam3}
\end{eqnarray}
The specifications (\ref{metr})-(\ref{gam3}) describe a manifold $M$ with 
the line element (\ref{ds2}), with {\it vanishing} nonmetricity and curvature,
\begin{equation}
Q_{\alpha\beta}=0, \quad\quad R_{\alpha}{}^{\beta}=0,\label{QR}
\end{equation}
but with nontrivial torsion,
\begin{eqnarray}
T^{\hat{1}}&=&{1\over f}\left({df\over dr}dr\wedge\vartheta^{\hat{1}} - 
2\,\vartheta^{\hat{2}}\wedge\vartheta^{\hat{3}}\right),\label{tor1}\\
T^{\hat{2}}&=&{1\over f}\left({df\over dr}dr\wedge\vartheta^{\hat{2}} - 
2\,\vartheta^{\hat{3}}\wedge\vartheta^{\hat{1}}\right),\label{tor2}\\
T^{\hat{3}}&=&{1\over f}\left({df\over dr}dr\wedge\vartheta^{\hat{3}} - 
2\,\vartheta^{\hat{1}}\wedge\vartheta^{\hat{2}}\right).\label{tor3}
\end{eqnarray}
For $2h=\pm {df/dr}$, the torsion is self- or anti-self-dual. 

Substituting this into (\ref{NiehYan}), we find for the Nieh-Yan four-form:
\begin{equation}
{1\over 2\ell^2}T_\alpha\wedge T^\alpha = -{6\over \ell^2}{df\over dr}f
\sin^2\psi\sin\theta\, dr\wedge d\psi\wedge d\theta \wedge d\phi.\label{TT}
\end{equation}
If integrated, this generically yields a nontrivial value for the invariant 
\begin{equation}
{1\over 2\ell^2}\int\limits_M T_\alpha\wedge T^\alpha. \label{topint}
\end{equation}
For the function $f(r)$ one could expect the instanton type of behavior
\begin{equation}
f={ar^2\over {r^2 + c^2}},\label{f}
\end{equation}
for example, where $a$ and $c$ are constants. However, since we do not 
discuss any gravitational field equations, it should be clear that (\ref{f}) 
is but one example.

%----------------------------------------------------------------------
\subsection{Zero connection gauge}

In a metric-affine spacetime, the connection can be ``gauged away'' at
any one point by a suitable local linear transformation of the frame.
However, in the teleparallel case under consideration, when the
curvature is trivial everywhere, see (\ref{QR}), one can always choose
a gauge in which the connection is vanishing {\it globally},
$\Gamma_\alpha{}^\beta=0$. A convenient way to demonstrate this is to
recall, following \cite{ChandiaZ}, that the hyperspherical coordinates
are related to the Cartesian coordinates $(x^1,x^2,x^3,x^4)$ via
\begin{eqnarray}
x^1&=&r\sin\psi\sin\theta\sin\phi,\\
x^2&=&r\sin\psi\sin\theta\cos\phi,\\
x^3&=&r\sin\psi\cos\theta,\\
x^4&=&r\cos\psi,
\end{eqnarray}
such that $r^2=(x^1)^2 +(x^2)^2 +(x^3)^2 +(x^4)^2$. Now define a {\it new} 
coframe:
\begin{eqnarray}
\vartheta^{\hat{4}}&=& hdr,\label{z1}\\
\vartheta^{\hat{1}}&=& {f\over r^2}\left(-x^2dx^1 + x^1dx^2 + x^4dx^3 -
x^3dx^4\right),\label{z2}\\
\vartheta^{\hat{2}}&=&{f\over r^2}\left(x^3dx^1 + x^4dx^2 - x^1dx^3 -
x^2dx^4\right),\label{z3}\\
\vartheta^{\hat{3}}&=&{f\over r^2}\left(x^4dx^1 - x^3dx^2 + x^2dx^3 -
x^1dx^4\right).\label{z4}
\end{eqnarray}
It is straightforward to check that (\ref{z1})-(\ref{z4}) and
\begin{equation}
g_{\alpha\beta}=\delta_{\alpha\beta}, \quad \Gamma_\alpha{}^\beta=0,
\end{equation}
define the same metric-affine spacetime with (\ref{ds2}) as line element, 
trivial nonmetricity and curvature (\ref{QR}), and a non-zero torsion which 
now equals the anholonomity 2-form:
\begin{equation}
T^\alpha=d\vartheta^\alpha.
\end{equation}
In components, we have again (\ref{tor1})-(\ref{tor3}), only the coframe
(\ref{cofr}) is replaced by (\ref{z1})-(\ref{z4}), and thus the
Nieh-Yan four-form (\ref{TT}) and the integral torsion invariant 
(\ref{topint}) are exactly the same. The coframe (\ref{z1})-(\ref{z4}) was 
discussed by Chandia and Zanelli \cite{ChandiaZ} who took, however, $f=1$ 
which, as is clear from (\ref{TT}), makes the Nieh-Yan four-form vanish 
identically.

%---------------------------------------------------------------------------
\subsection{Parallelizability and higher-dimensional example}

Since the Nieh-Yan form is constructed from the translational Chern-Simons
3-form,
\begin{equation}
C_{\rm TT}={1\over{2\ell^2}}\vartheta^\alpha\wedge T_\alpha =
-{1\over{2\ell^2}}\,{}^\ast\! A,\label{tcs}
\end{equation}
the new invariant (\ref{topint}) can be calculated as the integral
\begin{equation}
\int\limits_{S^3}C_{\rm TT}
\end{equation}
over the three-sphere of infinite radius. In our example, 
\begin{equation}
C_{\rm TT}=-{f^2(\infty)\over 2\ell^2}\,\underline{C}_{\rm TT},\quad\quad
\underline{C}_{\rm TT}=\underline{\vartheta}^a\wedge 
\underline{D}\underline{\vartheta}_a,
\end{equation}
where $\underline{\vartheta}^a$ is a three-coframe which lives on the
three-sphere $S^3$ of unit radius. Here, the underline denotes the geometrical 
objects on this $S^3$, and the Latin indices, from the beginning of the
alphabet, run over $a,b,...=1,2,3$. In our construction, we are thus naturally 
using the parallelizability of the three-sphere. Everything looks particularly
simple in the zero-connection gauge, when
\begin{equation}
\underline{C}_{\rm TT}=\underline{\vartheta}^a\wedge d\underline{\vartheta}_a
= -6\, \underline{\vartheta}^{\hat{1}}\wedge\underline{\vartheta}^{\hat{2}}
\wedge\underline{\vartheta}^{\hat{3}}.\label{tdt1}
\end{equation}
Thus  integration yields
\begin{equation}
\int\limits_{S^3}\underline{C}_{\rm TT} = -12\,\pi^2,                     
\end{equation}
since the volume of a sphere is 
\begin{equation}
{\rm vol}(S^{n-1})=\frac{n \pi^{n/2}}{\Gamma({n\over 2}+1)},\label{vol}
\end{equation}
that is, ${\rm vol}(S^3) = 2\pi^2$ and ${\rm vol}(S^7)= {1\over 3} \pi^4$.

These observations may be useful in discovering a further 
higher-dimensional example. As was noticed in \cite{ChandiaZ}, the 
next higher dimension, in which an analog exists of the Nieh-Yan form, is 8.
As is well known, among the spheres only $S^1$, $S^3$, and $S^7$ exhibit 
teleparallelism, i.e. are ``parallelizable''. Trautman \cite{Tra} suggested
to us to investigate the Dirac operator on the parallelizable non-trivial
spheres $S^3$ and $S^7$. One of us \cite{Budczies} did this and those 
techniques are also useful here in our context.

The anticipated generalization of the translational Chern-Simons term
is a seven-form
\begin{equation}
\underline{C}_{\rm TTTT}= \underline{\vartheta}^a\wedge 
d\underline{\vartheta}_a\wedge d\underline{\vartheta}^b 
\wedge d\underline{\vartheta}_b,\label{TTTT}
\end{equation}
and we need to integrate it over a seven-sphere. The indices evidently run
now from $a,b=1,\dots, 7$.

A convenient unified framework for treating all cases for $S^n$, $n=1,3,7$,  
is provided by the Clifford algebra approach. For $n=1,3,7$, the Clifford 
algebra over flat Euclidean space ${R}^n$ permits exactly an $n+1$-dimensional 
real representation. The corresponding generators are real $(n+1)\times (n+1)$
matrices $\underline{\gamma}_1, ..., \underline{\gamma}_n$ which satisfy the 
relations:
\begin{eqnarray}                                          
\underline{\gamma}_a\,\underline{\gamma}_b + \underline{\gamma}_b\,
\underline{\gamma}_a &=& -2\,\delta_{ab},\label{cliff} \\
\langle \underline{\gamma}_a v, \underline{\gamma}_a w\rangle  
&=& \langle v,w\rangle \qquad           
{\rm for\ all\ \ } v, w \in {R}^{n+1},        \label{ortho} \\
\underline{\gamma}_1 ... \underline{\gamma}_n &=&\pm 1\qquad\qquad 
({\rm for\ } n=3,7).\label{norm}
\end{eqnarray}
Here $\langle,\rangle$ denotes the Euclidean metric on ${R}^{n+1}$. Note that 
$\underline{\gamma}_a$ are different from the Dirac $\gamma$-matrices
which will appear in the next sections.

One can use $\underline{\gamma}_a$ to construct a natural $n$-bein for $S^n$ 
which we define as a submanifold 
$S^n:=\left\{p\in {R}^{n+1}\,\vert\, \langle p,p\rangle =1\right\}$.
First, using (\ref{cliff}), we rewrite (\ref{ortho}) as
\begin{equation} \label{skew}
\langle \underline{\gamma}_a v, w\rangle  = 
-\langle v, \underline{\gamma}_a w\rangle .
\end{equation}
Now, with $p\in S^n$, we define the vector fields 
\begin{equation}
{e_a}|_p := \underline{\gamma}_a p,\quad a=1,\dots n. 
\end{equation}
In view of the equations (\ref{cliff}) and (\ref{skew}), 
the $n+1$ vectors $(p, e_1|_p, ..., e_n|_p)$ form an orthonormal frame on 
${R}^{n+1}$. In particular, $e_1, ..., e_n$ is an orthonormal $n$-bein 
on $S^n$. The dual coframe is defined, for any $v$, by
\begin{equation}
\underline{\vartheta}_a(v) = \langle e_a, v\rangle .
\end{equation}
Now it is straightforward to find its exterior derivative:
\begin{equation}
d\underline{\vartheta}_a(e_b,e_c) = - \underline{\vartheta}_a([e_b, e_c]) 
= - 2 \, \langle \underline{\gamma}_a p, \underline{\gamma}_{[b}\,
\underline{\gamma}_{c]} p \rangle = 2\,\varphi_{abc},
\end{equation}
where we introduced the functions $\varphi_{abc}(p) := \langle 
\underline{\gamma}_{[a}\,\underline{\gamma}_b\,\underline{\gamma}_{c]}p,
p \rangle $ on ${R}^{n+1}$, which comprise a completely antisymmetric tensor.
Thus, for $n=1,3,7$, we find
\begin{equation}
\underline{\vartheta}^a \wedge d\underline{\vartheta}_a = \varphi_{abc}\; 
\underline{\vartheta}^a \wedge \underline{\vartheta}^b \wedge 
\underline{\vartheta}^c.\label{th^dth}
\end{equation}
For ${n=3}$ the functions $\varphi_{abc}$ are constant over $S^3$, namely 
$\varphi_{abc} = \pm\epsilon_{abc}$; then, with the minus sign, (\ref{th^dth}) 
reproduces (\ref{tdt1}).

For ${n=7}$, we have to work out
\begin{equation}
d\underline{\vartheta}^a\wedge d\underline{\vartheta}_a =\varphi^a{}_{bc}\,
\varphi_{ab'c'}\;\underline{\vartheta}^b\wedge\underline{\vartheta}^c\wedge
\underline{\vartheta}^{b'}\wedge\underline{\vartheta}^{c'}.\label{dth^dth}
\end{equation}
For $b\neq c,b'\neq c'$ we get:
\begin{eqnarray}
\label{phi*phi0}
(\varphi^a{}_{bc}\,\varphi_{ab'c'})(p)&=& \langle 
\underline{\gamma}_b\,\underline{\gamma}_c p,\underline{\gamma}^a p\rangle \;
\langle \underline{\gamma}_{b'}\,\underline{\gamma}_{c'}p,
\underline{\gamma}_a p\rangle = \langle \underline{\gamma}_{b'}\,
\underline{\gamma}_{c'}p,\langle \underline{\gamma}_b\,\underline{\gamma}_c p, 
\underline{\gamma}^a p\rangle  \underline{\gamma}_a p\rangle \nonumber\\
&=&\langle \underline{\gamma}_{b'}\,\underline{\gamma}_{c'}p,
\underline{\gamma}_b\,\underline{\gamma}_c p - 
\langle \underline{\gamma}_b\,\underline{\gamma}_c p,p\rangle p\rangle 
= \langle \underline{\gamma}_{b'}\,\underline{\gamma}_{c'} p,
\underline{\gamma}_b\,\underline{\gamma}_c p\rangle .
\end{eqnarray}
Let $b,c,b',c'$ be pairwise different and $\sigma$ be a permutation of 
$1, ...,7$, with $\sigma(4)=b,\sigma(5)=c,\sigma(6)=b',\sigma(7)=c'$. Then
(\ref{phi*phi0}) yields
\begin{eqnarray}
\label{phi*phi}
(\varphi^a{}_{bc}\,\varphi_{ab'c'})(p)&=& 
-\langle \underline{\gamma}_b\,\underline{\gamma}_c\,\underline{\gamma}_{b'}
\underline{\gamma}_{c'} p, p\rangle  
= -\langle \underline{\gamma}_{\sigma(4)}\,\underline{\gamma}_{\sigma(5)}\,
\underline{\gamma}_{\sigma(6)}\,\underline{\gamma}_{\sigma(7)}p, p\rangle  
\nonumber\\ &=& \mp{\rm sign}(\sigma)\,\langle \underline{\gamma}_{\sigma(1)}\,
\underline{\gamma}_{\sigma(2)}\,\underline{\gamma}_{\sigma(3)}p,p\rangle  
= \mp{\rm sign}(\sigma)\,\varphi_{\sigma(1)\,\sigma(2)\,\sigma(3)}(p),
\end{eqnarray}
where we used (\ref{norm}). Transvecting now (\ref{phi*phi0}) with
$\delta^{bb'}$, we find
\begin{equation}
  \varphi^{ab}{}_{c} \; \varphi_{abc'} = 6 \; \delta_{cc'}, \quad
  \varphi^{abc} \; \varphi_{abc} = 42. \label{phi*phi2}
\end{equation} 
For the seven-form (\ref{TTTT}), equations (\ref{th^dth})-(\ref{phi*phi2}) 
yield 
\begin{eqnarray}
\underline{C}_{\rm TTTT} &=& - \sum_{a,b,c,\sigma}{\rm sign}(\sigma)\;
\varphi_{abc}\,\varphi_{\sigma(1)\sigma(2)\sigma(3)}\;\underline{\vartheta}^a
\wedge\underline{\vartheta}^b\wedge\underline{\vartheta}^c\wedge
\underline{\vartheta}^{\sigma(4)}\wedge\underline{\vartheta}^{\sigma(5)}
\wedge\underline{\vartheta}^{\sigma(6)}\wedge\underline{\vartheta}^{\sigma(7)}
\nonumber\\ &=& \mp 4! \; \varphi_{abc}\, \varphi^{abc} \; 
\underline{\vartheta}^1\wedge\underline{\vartheta}^2\wedge
\underline{\vartheta}^3\wedge\underline{\vartheta}^4\wedge
\underline{\vartheta}^5\wedge\underline{\vartheta}^6\wedge
\underline{\vartheta}^7  \nonumber\\ 
&=& \mp 1008 \; 
\underline{\vartheta}^1\wedge\underline{\vartheta}^2\wedge
\underline{\vartheta}^3\wedge\underline{\vartheta}^4\wedge
\underline{\vartheta}^5\wedge\underline{\vartheta}^6\wedge
\underline{\vartheta}^7.\label{C_TTTT}
\end{eqnarray}
Two sign factors arise from rearranging the products of the coframe
one-forms: one is equal to the sign of the permutation
$(\sigma(1),\sigma(2),\sigma(3))$ of $(a,b,c)$ and, analogously,
another is equal to the sign of the permutation
$(a,b,c,\sigma(4),\sigma(5),\sigma(6),\sigma(7))$ of
$(1,2,3,4,5,6,7)$.  Their product yields ${\rm sign}(\sigma)$, and
thus all sign factors drop out.  Eventually, using (\ref{vol}), the
integral turns out to be
\begin{equation}
\int\limits_{S^7}\underline{C}_{\rm TTTT}
= \mp 24\times 42\times {\rm vol}(S^7) =\mp 336\,\pi^4.
\end{equation}

The 8-dimensional generalization of our example thus reads as follows: Add 
the radial coordinate $r$ to the seven angular coordinates, which parameterize 
$S^7$ and are used implicitly in $\underline{\vartheta}^a$. Then in the
gauge $\Gamma_\alpha{}^\beta =0$, we take the metric $g_{\alpha\beta}=
\delta_{\alpha\beta}$ and describe the coframe one-form by
\begin{equation}
\vartheta^{\hat 8}=hdr,\quad\quad \vartheta^{\hat 1}=f\underline{\vartheta}^1,
\quad\dots,\quad \vartheta^{\hat 7}=f\underline{\vartheta}^7,
\end{equation}
where $h=h(r),\,f=f(r)$ are two $SO(8)$-symmetric functions. 
A similar construction has been developed in \cite{englert} for the {\it 
Ricci--flat} Riemann--Cartan geometry on a seven-sphere. 

%----------------------------------------------------------------------
\section{Axial current in the Einstein--Cartan--Dirac theory}  

The Einstein--Cartan--Dirac (ECD) theory of a coupled gravitational and spin 
$1/2$ fermion field provide a classical (i.e., not quantized) understanding 
of the axial anomaly and establishes a link to the Nieh-Yan topological
invariant. The ECD-Lagrangian reads:
\begin{equation}
L={1\over 2\ell^2}R^{\alpha\beta}\wedge\eta_{\alpha\beta} + L_D,
\end{equation}
where the Dirac Lagrangian for the massless spinor field $\Psi$ is 
\begin{equation}
L_D=\frac{i}{2}\left\{\overline{\Psi}\,{}^\ast\gamma\wedge D\Psi
+\overline{D\Psi}\wedge{}^\ast\gamma\,\Psi\right\}.\label{10-4.10}
\end{equation}
As usually, $\eta^{\alpha\beta}:={}^\ast(\vartheta^\alpha\wedge
\vartheta^\beta)$ and $\eta^\alpha:={}^\ast\vartheta^\alpha$, overbar 
denotes the Dirac conjugate, and the constant Dirac matrices $\gamma^\alpha$
are entering via the Clifford-algebra valued exterior forms:
\begin{equation}
\gamma:=\gamma_\alpha\vartheta^\alpha\,,\qquad
{}^\ast\gamma=\gamma^\alpha\eta_\alpha\,.\label{10-4.2}
\end{equation}
The covariant exterior derivative $D$ for spinors is introduced by
\begin{equation}
  D\Psi=d\Psi+\frac{i}{4}\Gamma^{\alpha\beta}\wedge
{\sigma}_{\alpha\beta}\Psi,\quad\quad
  \overline{D\Psi}=d\overline{\Psi}-\frac{i}{4}\Gamma^{\alpha\beta}\wedge
\overline{\Psi}{\sigma}_{\alpha\beta}\, ,\label{10-4.9}
\end{equation}
where the Lorentz generator is represented by the components of the 
Clifford-algebra valued two-form
\begin{equation}\label{10-4.3}
\sigma:={i\over 2}\gamma\wedge\gamma = {1\over 2}\,{\sigma}_{\alpha\beta}
\,\vartheta^\alpha\wedge\vartheta^\beta\quad\Rightarrow\quad 
{\sigma}_{\alpha\beta}= \frac{i}{2}
(\gamma_\alpha\gamma_\beta-\gamma_\beta\gamma_\alpha)\,.
\end{equation}

The spin current of the Dirac field is given by the Hermitian three--form 
\begin{equation} 
\tau_{\alpha\beta}:={\partial L_{\rm D}\over\partial\Gamma^{\alpha\beta}} 
={1\over 8}\overline{\Psi}\left(\,^* \gamma\sigma_{\alpha\beta}+ 
\sigma_{\alpha\beta}\,^* \gamma\right)\Psi\, . 
\end{equation} 
{}From the anticommutation relations for the Dirac matrices we can infer that 
\begin{equation} 
\tau_{\alpha\beta}=\tau_{\alpha\beta\gamma}\,\eta^\gamma 
={i\over 4}\,\overline{\Psi}\gamma_{[\alpha}\gamma_\beta\gamma_{\gamma]}\Psi\, 
\eta^\gamma\,= -\frac{1}{4}\,\eta_{\alpha\beta\gamma\delta}\,\overline{\Psi} 
\gamma_5\gamma^{\delta}\Psi\eta^{\gamma}\,. \label{eq:spingamma} 
\end{equation} 
This implies that the components 
$\tau_{\alpha\beta\gamma}=\tau_{[\alpha\beta\gamma]}$ of the spin  
current are {\em totally antisymmetric}. 

The second field equation of EC-theory, i.e. the
algebraic  relation between torsion and spin, 
\begin{equation}
-{1\over 2} \eta_{\alpha\beta\gamma}\wedge T^{\gamma}=\ell^{2} 
\tau_{\alpha\beta}, \label{carspin}
\end{equation} 
can now be rewritten as
\begin{equation}
\vartheta^{\delta}\wedge T^{\gamma} = 
\frac{\ell^2}{2}\, \overline{\Psi}\gamma_5\gamma^{\delta} 
\Psi\eta^{\gamma}\,. \label{eq:etator} 
\end{equation}

The contraction of its free indices involves the {\em axial current}   
three-form of the Dirac field
\begin{equation}
j_5 := \overline{\Psi}\gamma_5\gamma^{\alpha} \Psi\eta_{\alpha}=
\overline{\Psi}\gamma_5\,^*\gamma \Psi . \label{eq:axial} 
\end{equation}
For the axial torsion one-form $A$, this implies $A=-(\ell^2/2) 
\overline{\Psi}\gamma_5\gamma\Psi$. Substituting it into the translational 
Chern-Simons term (\ref{tcs}), we find (cf. \cite{mie86})
\begin{equation}
C_{\rm TT} = \frac{1}{4}\,j_5 \,.\label{eq:shell} 
\end{equation}

{}From (\ref{NiehYan}) we thus find in ECD-theory, if a possible coupling 
to the Weyl covector is allowed for, 
\begin{equation}
dj_5 = 4 dC_{\rm TT} 
={2\over{\ell^2}} \left(T^\alpha\wedge
T_\alpha+R_{\alpha\beta}\wedge\vartheta^\alpha\wedge\vartheta^\beta\right)
-Q\wedge j_5\, . \label{eq:classan} 
\end{equation}

This result, cf. \cite{mie86,MMM96}, holds on the level of  first 
quantization. Since the Hamiltonian of the semi-classical Dirac field is 
not bounded from below, one has to go over to second quantization, where 
the divergence of the axial current picks up anomalous terms. The question 
is whether in the vacuum expectation value $<dj_5>$ similar torsion and 
Weyl covector terms emerge, besides the usual Pontrjagin term.

However, the Dirac equation cannot be extended to metric-affine gravity
 with its gauged 
$GL(4,R)$  group. Only an additional coupling to a Weyl covector $Q$ is 
admitted, which can surface in the Nieh-Yan term. Then the scale invariant 
covariant derivative $D := d +Q$ should differentiate the axial current.

%----------------------------------------------------------------------
\section{Heat kernel technique and axial anomaly} 
%%%%% EUCLIDean signature in this section

As a matter of fact, quantum anomalies both in the Riemannian and in the
Riemann-Cartan spacetimes (see \cite{Yuri1}) were calculated previously in a 
large number of papers using different methods (which prove to be all mutually 
consistent), see e.g. \cite{Odintsov} for a list of references and the more
recent papers \cite{Yajima,Wies96}. However, recently Chandia and Zanelli 
\cite{ChandiaZ} have questioned the completeness of the earlier calculations 
which all seem to demonstrate that the Nieh-Yan four-form is irrelevant to 
the axial anomaly. In this section we carefully reconsider the derivation of 
the axial anomaly for the massless Dirac field within the framework of the 
heat kernel technique. In particular, we will use the results of \cite{Yuri1} 
(though the reader should be careful because of the different notation). 

{}From (\ref{10-4.10}) we read off the Dirac operator $\Delta$ in  
Riemann-Cartan spacetime,
\begin{equation}
\eta\,\Delta\Psi := i\,{}^\ast\gamma\wedge
\left(D - {1\over 2}T\right)\Psi,\label{dirac}
\end{equation}
where, as usual, $T=e_\alpha\rfloor T^\alpha$ is the torsion trace one-form. 
The square of $\Delta$ is a self-adjoint second order differential operator
${\cal D}^2:=\Delta\Delta^\dagger$,
\begin{equation}
\eta\,{\cal D}^2 = - (D{}^\ast\! D + 2\,S\wedge{}^\ast\! D + X),\label{dd}
\end{equation}
where, as follows from (\ref{dirac}), the one-form $S$ is defined by
\begin{equation}
S:=\,{i\over 2}\,\vartheta_\alpha{}^\ast(^\ast\sigma\wedge T^\alpha),\label{S}
\end{equation}%%%%%%Sign for Euclid
and the four-form $X$ reads:
\begin{equation}
X:={1\over 4}\,{}^\ast\sigma\wedge R_{\alpha\beta}\,\sigma^{\alpha\beta} +
{1\over 2}\left(-d\,{}^\ast T + i\,{}^\ast\sigma\wedge dT - {1\over 2}
T\wedge{}^\ast T + i\,(e_\alpha\rfloor{}^\ast\sigma)\wedge 
T^\alpha\wedge T\right).\label{X}
\end{equation}

Among the operator functions which can be constructed from ${\cal D}^2$,
the expression $\exp(-t{\cal D}^2),\ t>0$, is of particular importance. It 
is formally defined by its kernel {\it four-form} $K(t,x,y,{\cal D}^2)$:
\begin{equation}
\exp(-t{\cal D}^2)\,\psi(x):=\int\,K(t,x,y,{\cal D}^2)\,\psi(y).
\end{equation}
Here $x,y$ are points of the spacetime manifold $M$ and the integral is 
over $M$ parameterized by the local coordinates $y$. By construction, $K$ 
satisfies the heat equation
\begin{equation}
{\partial \over \partial t}K(t,x,y,{\cal D}^2) + 
{\cal D}^2\,K(t,x,y,{\cal D}^2)=0,\quad\quad 
K(0,x,y,{\cal D}^2)=\delta(x,y).
\end{equation}
For small $t\rightarrow +0$, there exists an asymptotic expansion,
\begin{equation}
K(t,x,x,{\cal D}^2)=\sum_{n=0}^\infty t^{{n\over 2} -2}(4\pi)^{-2} 
K_n(x,{\cal D}^2),\label{exp}
\end{equation}
where the coefficients $K_n(x,{\cal D}^2), n=0,1,\dots$ are the four-forms
on spacetime which are completely determined by the form of the positive 
second-order differential operator ${\cal D}^2$, namely by the exterior 
forms $S$ and $X$. In \cite{Yuri1} the general expression for these 
coefficients was derived for a four-dimensional Riemann-Cartan manifold $M$ 
and an arbitrary operator of the form (\ref{dd}) acting on the sections of 
a bundle with internal curvature two-form $F$. For odd $n=1,3,\dots$ the
coefficients are zero, while the first nontrivial terms read
\begin{eqnarray}
K_0&=&\hbox{\boldmath $1$}\eta,\label{K0}\\
K_2&=&Z + {1\over 6}\widetilde{R}\eta,\label{K2}\\
K_4&=&{1\over 3}({\rm com}) + {1\over 6}\widetilde{D}{}^\ast\!\widetilde{D}
\left({}^\ast Z + {1\over 5}\widetilde{R}\right) +
{1\over 2}Z\,{}^\ast Z + {1\over 6}\widetilde{R}Z 
+ {1\over 6}{}^\ast Y\wedge Y\nonumber\\ 
&&+ {1\over 180}\left(2{}^{\ast (1)}\widetilde{R}^{\alpha\beta}\wedge
{}^{(1)}\widetilde{R}_{\alpha\beta}
+{}^{\ast (4)}\widetilde{R}^{\alpha\beta}\wedge
{}^{(4)}\widetilde{R}_{\alpha\beta}
+29{}^{\ast (6)}\widetilde{R}^{\alpha\beta}\wedge
{}^{(6)}\widetilde{R}_{\alpha\beta}\right).\label{K4}
\end{eqnarray}%%%%%%% for {}^\ast Z Euclid sign is plus (as above)
Here $\hbox{\boldmath $1$}$ is the unity matrix,
\begin{equation}
Z:=X - d{}^\ast\! S - S\wedge{}^\ast S,\quad\quad
Y:=F + dS + S\wedge S,\label{ZY}
\end{equation}
the tildes denote purely Riemannian 
objects and operators, and the superscripts label the irreducible pieces of 
the Riemannian curvature (1st represents the Weyl piece, 4th -- the traceless
Ricci, and 6th -- the curvature scalar $\widetilde{R}:=e_\alpha\rfloor 
e_\beta\rfloor\widetilde{R}^{\alpha\beta}$). Finally, $({\rm com})$ is an 
inessential term which satisfies ${\rm Tr}({\rm com})={\rm Tr}(\gamma_5
({\rm com}))=0$. 

It is well known that the axial anomaly is closely related to the 
Atiyah-Singer index theorem \cite{Atiyah}. Technically, this leads to the 
calculation of the trace
\begin{equation}
{\rm Tr}\left(\gamma_5\,e^{-t{\cal D}^2}\right),\quad t\longrightarrow +0,
\end{equation}
for example, using the Fujikawa method. In view of the heat kernel 
expansion (\ref{exp}), this problem eventually reduces to the computation 
of ${\rm Tr}(\gamma_5\,K_n$). 

For the Dirac operator, evidently 
\begin{equation}
{\rm Tr}(\gamma_5\,K_0)=0,
\end{equation} 
while the result for $n=4$ was given explicitly in \cite{Yuri1} in terms 
of the Riemannian Pontrjagin four-form,
\begin{equation}
{\rm Tr}(\gamma_5\,K_4)={1\over 12}\left(\widetilde{R}^{\alpha\beta}\wedge
\widetilde{R}_{\alpha\beta} + {1\over 2}dA\wedge dA + d{\cal K}\right),
\end{equation}
where, to recall, $A$ is the axial torsion one-form, and in the boundary
term the three-form ${\cal K}$ is constructed from $A$. 

Finally, substituting (\ref{S}) and (\ref{X}) into (\ref{ZY})
and (\ref{K2}), we find 
\begin{eqnarray}
{\rm Tr}(\gamma_5\,K_2)&=& {\rm Tr}(\gamma_5\,Z)={\rm Tr}(\gamma_5\,[X -
S\wedge{}^\ast S])\nonumber\\ &=&{1\over 4}{\rm Tr}(\gamma_5\,
{}^\ast\sigma\,\sigma^{\alpha\beta})\wedge\left(R_{\alpha\beta} + 
{1\over 2}T^\gamma\,{}^\ast(\eta_{\alpha\beta}\wedge T_\gamma)\right)
\nonumber\\ &=&-\,\left(R_{\alpha\beta}\wedge\vartheta^\alpha\wedge
\vartheta^\beta + T^{\alpha}\wedge T_\alpha\right),\label{trk2}
\end{eqnarray}
where we used the identity $\eta_{\alpha\beta}\wedge\Phi = -\eta\, e_\alpha
\rfloor e_\beta\rfloor\Phi$ valid for any two-form $\Phi$, and the trace
\begin{equation}
{1\over 4}{\rm Tr}(\gamma_5\,{}^\ast\sigma\,\sigma^{\alpha\beta})= -\,
\vartheta^\alpha\wedge\vartheta^\beta.
\end{equation}
[Note that a sign factor (and/or $i$) may appear which depends on the
signature and the representation of the Dirac matrices; in this section
we work with the Euclidean version.]

It is not necessary to compute further terms ${\rm Tr}(\gamma_5\,K_n$) for
$n>4$ because they are multiplied by the positive powers of $t$ and drop
out when $t\rightarrow 0$.

Thus, using the heat kernel technique and the earlier results on the 
spectral geometry of the Riemann-Cartan spacetime (\ref{K0})-(\ref{K4}),
we are able to confirm the proposition of Chandia and Zanelli that the 
Nieh-Yan four-form can, indeed, show up in the study of the chiral anomaly.
This is demonstrated explicitly by (\ref{trk2}). 

There is a problem, though: 
the Nieh-Yan four-form enters the heat kernel expansion with the necessarily 
divergent coefficient ${1\over t}$, with $t\rightarrow 0$. Formal 
regularization of the axial anomaly in the zeta-function or dimensional 
schemes leads to the complete subtraction of this term which explains 
why it was never reported in the literature. However, in the Fujikawa 
method, used in \cite{ChandiaZ}, this term survives and is proportional 
to the regulator mass square $M^2\rightarrow\infty$.

Chandia and Zanelli proposed to absorb this divergence by a proper 
rescaling of the coframe. We can comment here that such a renormalization
should be only possible in a conformally invariant gravity theory with
massless fermions. Then, the usual Einstein-Cartan theory is
not applicable. 

%----------------------------------------------------------------------
%\section{Conclusion} 

\section*{Acknowledgments} 

We thank {\it Jorge Zanelli} for useful discussions and for providing
a preliminary draft of Chandia's and his paper. We also acknowledge
interesting discussions with Alfredo Mac\'{\i}as.  This work was
partially supported by KFA--Conacyt Grant No. E130--2924.  One of us
(EWM) acknowledges the support by the short--term fellowship 961 616
015 6 of the German Academic Exchange Service (DAAD), Bonn.  For (YNO)
this work was supported by the Deutsche Forschungsgemeinschaft (Bonn)
under contract He 528/17-2.

%\begin{thebibliography}{999}%for Latex 
 
\end{document}